\documentclass[runningheads]{llncs}
\usepackage[T1]{fontenc}
\usepackage{graphicx}
\usepackage{paralist}
\usepackage{orcidlink}
\usepackage{array}
\usepackage{booktabs}
\newcolumntype{L}[1]{>{\raggedright\arraybackslash}p{#1}}
\begin{document}
\title{ContinuumBench: Benchmarking Joint Autoscaling and Placement Across Evaluation Regimes in the Cloud-Edge Continuum}
\titlerunning{ContinuumBench}
%
\author{Lanpei Li\inst{1,2}~\orcidlink{0009-0005-4370-1020}\and
Antonino Vaccarella\inst{1,2}~\orcidlink{0009-0004-0428-540X}\and
Vincenzo Lomonaco\inst{3}~\orcidlink{0000-0001-8308-6599}\and
Massimo Coppola\inst{1}~\orcidlink{0000-0002-7937-4157}}
\authorrunning{L. Li et al.}
%

\institute{Institute of Information Science and Technologies ``Alessandro Faedo'' (ISTI), National Research Council of Italy (CNR), 56124 Pisa, Italy
\and
Department of Computer Science, University of Pisa, 56127 Pisa, Italy
\and
Department of AI, Data and Decision Sciences, \\ LUISS University, 00197 Rome, Italy \\
\email{\{lanpei.li, massimo.coppola\}@isti.cnr.it\\
antoninovaccarella@cnr.it, vlomonaco@luiss.it}}

\maketitle              
\begin{abstract}
Cloud–edge controllers coordinate service placement, replica scaling, and resource pre-warming to keep end-to-end latency within application deadlines. But evaluations often obscure the source of a reported gain: placement and scaling are studied separately; workload, connectivity, and calibration assumptions remain implicit; and metrics over completed tasks hide unfinished work. We present \textit{ContinuumBench}, a benchmark that controls these factors. Its \emph{completion-aware accounting} treats late, unfinished, and discarded tasks as deadline misses. A common protocol compares placement-only and scale-capable controllers under declared regimes and stressors. Built on the \textit{ECLYPSE} simulator, ContinuumBench adds arrivals, worker elasticity, intermittent transport, buffering, and failures to close the control loop. We evaluate nine controllers across four scenarios and two regimes. The studied regimes are capacity-bound: elastic capacity, not placement sophistication, drives completion, and once capacity suffices, the choice of autoscaling policy decides how much of that work arrives on time. Placement re-planning has no measurable effect without relocation, while cost-free migration defines the observed exception. Consequently, scale-capable controllers approach an over-provisioned reference while placement-only controllers degrade with load; and placement quality separates controllers only once capacity is exhausted. Finally, the accounting choice itself changes the reported result: completion-only and completion-aware scoring can rank controllers differently.

\end{abstract}

\renewcommand{\thefootnote}{\roman{footnote}}
\footnotetext[0]{
This version of the contribution has been accepted for publication, after peer review but is not the Version of Record and does not reflect post-acceptance improvements, or any corrections. The work was presented at the 6th workshop on Flexible Resource and Application Management on the Edge (FRAME) 2026, co-located with the 32nd International European Conference on Parallel and Distributed Computing -- Euro-Par 2026. The Version of Record will appear in the workshop proceedings volume(s) of Euro-Par 2026. Use of this Accepted Version is subject to the publisher's Accepted Manuscript terms of use \url{https://www.springernature.com/gp/open-research/policies/accepted-manuscript-terms}.
}
\renewcommand{\thefootnote}{\arabic{footnote}}
\setcounter{footnote}{0}%

\section{Introduction}
Cloud-edge resource management coordinates placement, scaling, offloading, and buffering across heterogeneous Internet of Things (IoT), edge, and cloud resources~\cite{ullah2023cloudtothings}. Latency-sensitive applications, including vehicular offloading, split AI inference, industrial monitoring, and intermittently connected remote sensing, can be represented as directed acyclic graphs (DAGs) of stages with latency budgets, resource demands, and placement constraints. A \emph{controller} selects each stage's location and number of active \emph{workers}, i.e., parallel replicas. An evaluation fixes a service graph, workload, network, and scoring method. Because these choices can determine the apparent winner, a measured advantage may reflect the evaluation rather than the controller. We call this \emph{evaluation dependence}~\cite{kasture2016tailbench,gan2019deathstarbench}.

Two continuum properties require controlled evaluation. First, placement and scaling interact: a good placement cannot offset insufficient burst capacity, whereas scaling without placement awareness can waste scarce edge resources. Prior work therefore couples these decisions~\cite{jointautoscalingplacement2021,cheng2023proscale}; we test whether coupling is necessary for on-time delivery. Second, intermittent connectivity can make a structurally feasible path unavailable at runtime~\cite{kodheli2021satcomsurvey}. Evaluations must therefore track buffered and unfinished tasks as contacts open and close.

We study \emph{when control sophistication improves cloud-edge resource management and how delivered work must be measured}. We present \textit{ContinuumBench},\footnote{Code and scenarios are publicly available (MIT license): \url{https://github.com/lilanpei/ContinuumBench}.} a benchmark that compares autoscaling and placement controllers across controlled regimes. It builds on the \textit{ECLYPSE} cloud-edge continuum simulator~\cite{massa2026eclypse} and separates runtime dynamics from substrate feasibility. We make four contributions:

\begin{itemize}
    \item \textbf{A control layer over ECLYPSE.} ContinuumBench adds an epoch-based loop to ECLYPSE's fixed substrate. ECLYPSE audits each action before execution, preventing infeasible states. Nine controllers---placement heuristics, an exact optimizer, and the Kubernetes Horizontal Pod Autoscaler (HPA) and Event-Driven Autoscaling (KEDA)---use a common action representation and observation set.

    \item \textbf{An evaluation method.} \emph{Completion-aware accounting} treats late, unfinished, and discarded tasks as deadline misses, so work that never finishes stays in the score, and it makes explicit how much the accounting choice alone changes the reported ranking.

    \item \textbf{An evaluation protocol.} Controlled ablations fix all variables except the factor under study; declared regimes and stressors make every reported difference auditable. Runtime models cover arrivals, elastic workers, intermittent transport, buffering, and failures, with per-epoch traces.

    \item \textbf{A design question put to controlled test.} We ask which control lever governs on-time delivery. Three ablations isolate one lever each---the re-planning schedule, elastic capacity under rising load, and placement quality under a placement bottleneck---while holding the evaluation fixed. Section~\ref{sec:results} reports what each yields.
\end{itemize}

\section{Related Work}
\label{sec:related}
\paragraph{\textbf{Continuum resource management.}}
Surveys like \cite{ullah2023cloudtothings} explore the issues of placement, offloading, scheduling and orchestration across the continuum. Moving closer to our study, joint autoscaling-and-placement methods optimize both decisions on the premise that placement is ineffective without sufficient worker capacity~\cite{jointautoscalingplacement2021}, whereas ProScale forecasts per-microservice workload and jointly sets instance counts, placement, and offloading~\cite{cheng2023proscale}. Both report improvements under study-specific graphs, traces, network conditions, and metrics, so neither separates gains due to the control lever from gains due to the evaluation setup. ContinuumBench holds the workload, regime, stressors, and completion-aware metric fixed while comparing placement only, scaling with fixed placement, and joint placement and scaling.

\paragraph{\textbf{Continuum simulators.}}
Continuum simulators model the substrate on which a controller operates. iFogSim models topology, resources, and energy~\cite{gupta2017ifogsim}; faas-sim drives serverless-edge experiments with measured traces~\cite{raith2023faassim}; and ECLYPSE models placement feasibility, residual resources, and routing~\cite{massa2026eclypse}. These simulators leave the control loop and evaluation protocol to the user: which controller acts, under which conditions, and how delivered work is scored. ContinuumBench adds this layer to ECLYPSE.

\paragraph{\textbf{Systems benchmarking.}}
Systems benchmarks standardize methodology. TailBench defines tail-latency measurements~\cite{kasture2016tailbench}, and DeathStarBench provides realistic microservice graphs~\cite{gan2019deathstarbench}. Neither targets placement or autoscaling control, and both report latency over completed requests. That assumption fails when controller decisions affect whether requests complete. ContinuumBench instead varies the controller under declared regimes and stressors, and counts unfinished and discarded tasks as deadline misses.

\begin{figure}[t]
    \centering
    \includegraphics[width=\linewidth,trim={20 14 20 14}]{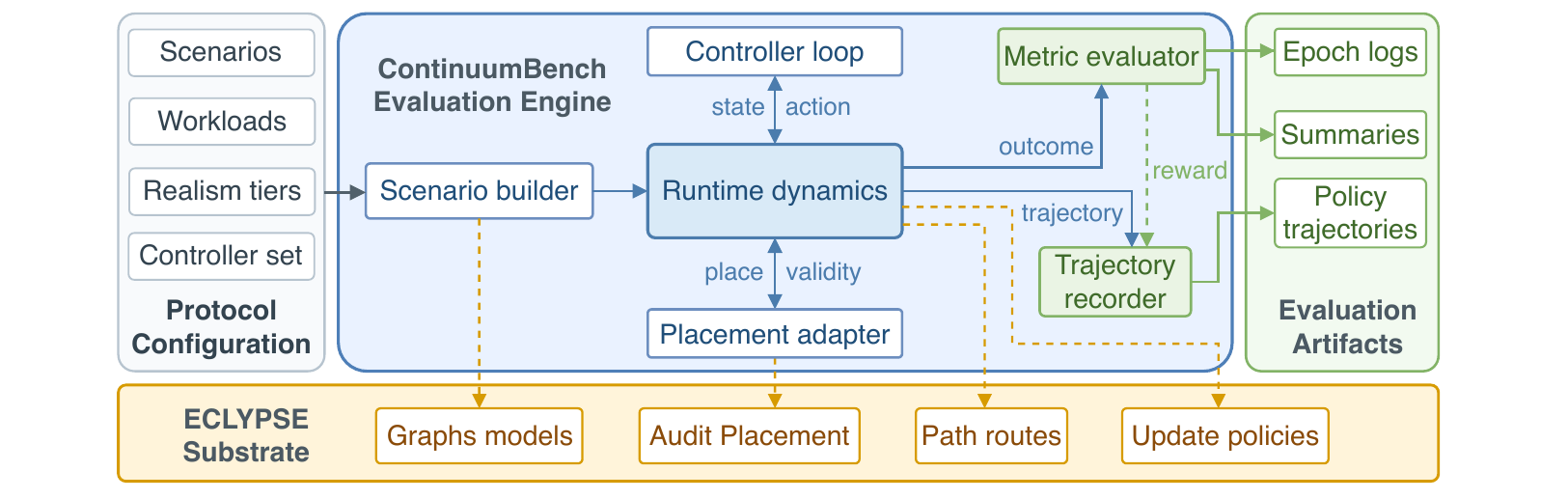}
    \caption{ContinuumBench extends the ECLYPSE substrate with workload arrivals, time-varying contacts, and worker scaling. ECLYPSE provides placement-feasibility checks, residual-resource state, and routing.}
    \label{fig:architecture}
\end{figure}

\section{ContinuumBench Architecture}
\label{sec:architecture}
A ContinuumBench run is defined by six configurable components: an application graph, an infrastructure graph, a workload process, a controller, an execution regime, and an analysis configuration.

The infrastructure graph is $G_I=(N,L)$. Each finite-capacity node $n\in N$ belongs to a tier $\tau(n)\in T=\{\mathrm{IoT},\mathrm{edge},\mathrm{cloud},\mathrm{space}\}$ and has CPU/RAM capacity $c_n$. Each link $e \in L$ has a latency, a bandwidth, and optional time-varying availability. Space--edge links are intermittent and follow satellite contact windows.

The application graph $G_A=(S,E_A)$ is a DAG of stages $s\in S$. We classify each stage by its dataflow role as a \textit{source}, \textit{transform} (intermediate processing, splits and joins), or \textit{sink}.

Each stage specifies a tier-anchoring set $\mathcal{T}_s\subseteq T$ on which it may be placed\footnote{Anchored stages are pinned to their origin tier; \textit{placeable} ones are freely deployable.}, a resource demand $r_s$, a per-task service time $\mu_s$, an optional Service Level Objective (SLO) deadline $\delta_s$, and bounds $[w_s^{\min},w_s^{\max}]$ on its number of active workers. All scenarios use one end-to-end deadline at the sink, measured from task creation at the source to task completion at the sink.

\paragraph{\textbf{Problem setting.}} We study the service-placement problem of~\cite{taleb2025survey} coupled with elastic scaling. At each fixed-length control epoch, the controller selects a joint action $a(t)=(x,w)$: the binary variable $x_{s,n}$ records whether stage $s$ is placed on node $n$, while $w_s\in[w_s^{\min},w_s^{\max}]$ sets its active worker count. Routing is fixed across controllers. Each source assigns a priority class to every task; a fixed triage rule sends high-priority tasks over the latency-sensitive path and all other tasks to the elastic pool; and each stage dispatches ready tasks round-robin among reachable workers. Every action must satisfy node capacity (placed demand $\le c_n$), tier anchoring ($x_{s,n}=1$ only if $\tau(n)\in\mathcal{T}_s$; only source stages are anchored here), and the ECLYPSE feasibility audit. Contacts and failures change transport availability but not these structural constraints. Subject to them, the controller maximizes completion-aware on-time delivery (Section~\ref{sec:metrics}).

\paragraph{\textbf{Control mechanism.}} At each epoch, all controllers observe per-stage queue depth---tasks awaiting a worker, plus tasks buffered awaiting transport at elastic pools---oldest-task age, time-to-live (TTL) pressure, and worker-pool state (active, starting, and ready counts). TTL pressure is the maximum age-to-TTL ratio among tasks held in store-and-forward buffers, and approaches $1$ as the most at-risk task nears expiry.

Scaling changes only worker activation. Each elastic stage has $w_s^{\max}$ pre-provisioned workers; the controller activates a count within $[w_s^{\min},w_s^{\max}]$, and each newly activated worker incurs a configurable startup delay. Active throughput is $w_s/\mu_s$ tasks per second; the bounded worker pool defines the capacity the controller can activate at runtime.

A placement action proposes a complete mapping and takes effect only after the ECLYPSE audit. The heuristics mirror incremental production autoscalers: they place newly activated workers but do not relocate existing ones, so only the exact optimizer of Section~\ref{sec:experimental-protocol} performs \emph{migration}. Stage-keyed queues and buffers preserve queued work when placement changes redirect subsequent processing. Migration downtime is zero throughout, testing relocation at its best: if on-time delivery still fails to improve, overhead cannot be the reason.

ContinuumBench implements this loop on the ECLYPSE substrate (Figure~\ref{fig:architecture}), which supplies the graphs, placement and resource checks, path lookup, and transfer-cost estimation~\cite{massa2026eclypse}; an action failing the audit is rejected whole, retaining the previous placement and worker counts. Each control epoch executes three steps:
\begin{inparaenum}[\bf I.]
    \item Contact schedules and failure policies update the time-varying link and node state.
    \item The controller updates the placement mapping, the worker counts, or both.
    \item ContinuumBench flushes buffered work when contacts permit, admits new arrivals, routes and processes ready tasks, records completed or failed work, and aggregates the resulting state into per-epoch summaries.
\end{inparaenum} 

The loop separates \emph{structural feasibility}, whether ECLYPSE can legally map stages to nodes, from \emph{transport availability}, whether the required links are reachable at a given time. An unavailable link or transient failure therefore delays, buffers, or drops a task without invalidating the active mapping, which stays fixed while buffered work awaits store-and-forward delivery.

\section{Scenarios and Workloads}
\label{sec:scenarios}
\begin{figure}[t]
    \centering
    \includegraphics[width=\linewidth, trim={20 17 20 17}]{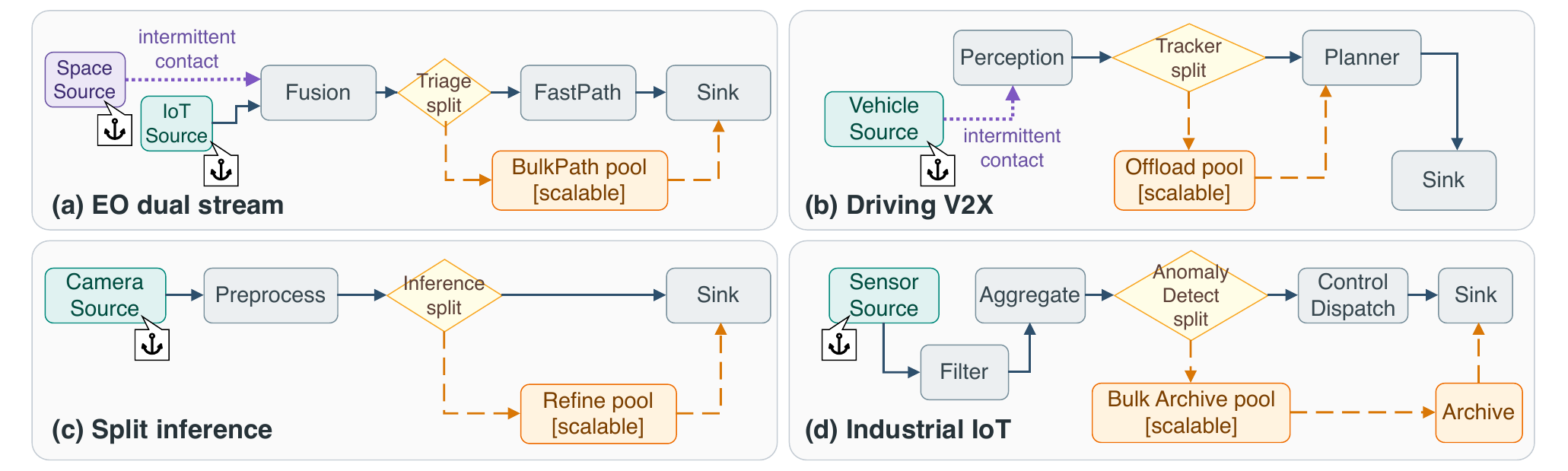}
    \caption{Scenario workflows evaluated in ContinuumBench. Solid arrows denote latency-sensitive, high-priority paths; orange dashed arrows denote elastic paths controlled by placement and scaling; purple dotted arrows denote intermittent-contact boundaries. Anchor symbols identify stages with fixed placement.}
    \label{fig:scenario-workflows}
\end{figure}

ContinuumBench defines four scenario families (Figure~\ref{fig:scenario-workflows}), each combining an application DAG, infrastructure graph, and workload process. All use a source--transform--sink structure but isolate different cloud-edge resource-management mechanisms.

\begin{description}
    \item[\textbf{EO dual stream}] Earth-observation (EO) sensing over intermittent links, with anchored space/IoT sources, a fast path, and an elastic bulk path; stresses contact windows and store-and-forward buffering~\cite{kodheli2021satcomsurvey}.
    \item[\textbf{Driving V2X}] Low-latency vehicle-to-everything (V2X) offloading with an anchored vehicle source, a critical planner path, and an elastic offload pool; stresses source-to-processing reachability~\cite{Liu2020vehicularedgesurvey}.
    \item[\textbf{Split inference}] Split AI serving with an anchored camera source, an edge-hosted early-exit path, and an optional elastic pool of heavier refinement workers; stresses the placement--worker-activation tradeoff~\cite{teerapittayanon2017distributed}.
    \item[\textbf{Industrial IoT}] Plant monitoring with an anchored sensor source, a deadline-sensitive path for rare anomalies, and an elastic archival path for bulk telemetry; stresses mixed-criticality SLO separation~\cite{xu2018industrialsurvey}.
\end{description}

Each scenario draws arrivals from a seeded Poisson process with piecewise-constant rates, giving repeatable low-load, burst, and recovery phases; contact-trace variants apply recorded traces only at intermittent boundaries.

The default infrastructure is a hand-built tiered graph (IoT$\to$\allowbreak edge$\to$\allowbreak cloud, space$\to$\allowbreak edge). The edge--cloud and space--edge tier pairs are fully connected, each IoT node attaches round-robin to a single edge node, and the only intra-tier links mesh the edge tier over a LAN (1\,ms, 10\,Gbps). Node capacities $c_n$ (CPU cores, RAM in GB) are $(128,256)$ for cloud, $(24,64)$ for edge, $(8,16)$ for space, and $(4,8)$ for IoT. Inter-tier links default to 20\,ms/1\,Gbps for edge--cloud and 6\,ms/120\,Mbps for IoT--edge; the space--edge link has 220\,ms latency and provides 45\,Mbps during contact. These configurable values define R1.

Stage demands $r_s$ range from 0.5 cores at source stages to 8 at the heaviest inference stage, and placeable stages may share a node when capacity permits. Each run contains 120 workload epochs followed by 60 arrival-free \emph{drain} epochs of 1\,s, the drain letting in-flight work finish before metrics are computed (Table~\ref{tab:scenario-scales}).

\begin{table}[t]
    \centering
    \caption{Scenario scale and per-regime workload. \emph{Stages} counts DAG stages with elastic pools at maximum size; \emph{Nodes} reports the space/IoT/edge/cloud split; \emph{Wrk.} is the worker-pool cap; and \emph{Tasks} is the expected number of arrivals over 120 workload epochs (R1/R2 where they differ). R2 adds trace-driven contact windows for EO and Driving V2X. The DAG, topology, capacities, and stage demands are identical across regimes.}
    \label{tab:scenario-scales}
    \footnotesize
    \setlength{\tabcolsep}{4pt}
    \renewcommand{\arraystretch}{1.0}
    \begin{tabular}{@{}lcccccc@{}}
        \toprule
        \textbf{Scenario} & \textbf{Stages} & \textbf{Nodes} & \textbf{Wrk.} & \textbf{$\approx$Tasks} & \textbf{$\delta$\,(s)} & \textbf{R2 $\mu_s$} \\
        \midrule
        EO dual stream & 10 & 6 (1/2/2/1) & 4 & 490/464 & 100 & unchanged \\
        Driving V2X & 8 & 6 (0/3/2/1) & 3 & 380/348 & 90 & unchanged \\
        Split inference & 6 & 4 (0/1/2/1) & 2 & 320/380 & 70 & 0.16--1.02\,s \\
        Industrial IoT & 10 & 5 (0/2/2/1) & 3 & 200 & 10 & 0.12--1.00\,s \\
        \bottomrule
    \end{tabular}
\end{table}

\section{Evaluation Regimes, Stressors and Metrics}
\label{sec:regimes}
ContinuumBench declares the regime and stressors for every run. Controlled ablations fix the application DAG, action space, feasibility constraints, and analysis pipeline and vary one stated factor. Cross-regime comparisons are descriptive when multiple declared inputs differ. Seeded draws make arrival counts, priority classes, and failure events exactly repeatable.

\textbf{R1} is the seeded synthetic baseline defined in Section~\ref{sec:scenarios}. \textbf{R2} changes up to four calibrated inputs, as listed in Table~\ref{tab:scenario-scales}: per-stage service times $\mu_s$, contact windows (periodic $\to$ trace-driven), arrival-rate schedules, and one \emph{inter-tier} link per scenario; all other links retain R1 values. The calibrated inter-tier link follows data availability, not expected placement. Following~\cite{raith2023faassim,horvath2024sealcc}, we use representative space--edge and edge--cloud profiles rather than reproduce measurements; IoT--edge remains synthetic. For example, Split inference changes edge--cloud from $20$\,ms/1\,Gbps to $34$\,ms/160\,Mbps.

We vary one \emph{stressor} at a time on top of either regime: arrival load, contact severity, or node unavailability. The last independently fails each node with probability $p\in\{0.03,0.08\}$ per epoch; failures last one epoch and are redrawn.

\paragraph{\textbf{Metrics and analysis pipeline.}}
\label{sec:metrics}
ContinuumBench reports delivered, pending, and failed work, latency, and SLO violations at epoch and run granularity. Let $G$ be generated tasks, $C$ completed tasks, $D_b$ drops from full store-and-forward buffers, $X_b$ buffer expirations after the 90--300\,s per-stage time-to-live, and $F_n$ node-failure losses. Failed work is $F=D_b+X_b+F_n$, and pending work is $P=\max(0,\,G-C-F)$. Completion, pending, and failure rates are $C/G$, $P/G$, and $F/G$.

For each completed task $i$, latency is $\ell_i=t_i^{\mathrm{complete}}-t_i^{\mathrm{created}}$; we report p95. This excludes unfinished work and can favor a controller that leaves late tasks incomplete, so we pair it with \emph{completion-aware accounting}.

A task's \emph{deadline instant} is its creation time plus the end-to-end deadline $\delta$ (Table~\ref{tab:scenario-scales}). At run end, we consider only tasks whose deadline has elapsed. Among them, $A$ completed on time, $B$ completed late, and $U$ did not complete---they remain in flight or were discarded, expired, or lost to a node failure. Thus, $E_{\mathrm{slo}}=A+B+U$, $V_{\mathrm{slo}}=B+U$, and the SLO-violation rate is $(B+U)/(A+B+U)$; where a single deadline-sensitive figure is needed, we report \emph{on-time delivery} $A/G$. Late, unfinished, and discarded eligible tasks all count as misses, so a controller cannot score better simply because eligible work never finished. Tasks with future deadlines remain outside the score; the 60-epoch drain reduces this boundary set. Treating unfinished work as missed follows firm real-time deadline metrics~\cite{buttazzo2011hard} and completed-only scoring concerns in LLM serving~\cite{wang2024revisitingslo}.

For each scenario--regime pair, we select the best \emph{placement-only} controller, which changes only the mapping, and the best \emph{scale-capable} controller, which also changes worker counts. The completion-aware score is $C/G \;-\; 0.5\,P/G \;-\; V_{\mathrm{slo}}/E_{\mathrm{slo}} \;-\; 0.01\,\ell^{\mathrm{p95}} \;-\; 0.001\,m$, where $\ell^{\mathrm{p95}}$ is p95 latency in seconds and $m$ is migrations. Latency and migration are secondary penalties. Perturbing all four weights over an 81-point grid changes the selected representative in one of the eight scenario--regime contexts, where both candidates are scale-capable; the family-level comparison never changes.

\section{Experimental Setting}
\label{sec:experimental-protocol}
We evaluate four placement-only and five scale-capable controllers (Table~\ref{tab:controllers}), including the deployed Kubernetes policies HPA and KEDA and a mixed-integer linear programming (MILP) controller bounding the headroom from placement alone. Scale-capable controllers place newly activated workers with the min-latency strategy, an earliest-finish-time-style greedy rule~\cite{topcuoglu2002heft}; Meta-rule picks its own, and the load sweep reuses min-latency as a fixed-capacity placement-only reference. Placement is recomputed every second epoch, and after each scale event for the joint heuristic; startup and migration delays are zero throughout.

\begin{table}[t]
    \centering
    \caption{Controllers evaluated in ContinuumBench. All controllers receive the same observations (Section~\ref{sec:architecture}); \emph{worker} is what HPA and KEDA call a replica. Parameter ranges cover the eight scenario--regime configurations, with one fixed value used in each run.}
    \label{tab:controllers}
    \footnotesize
    \setlength{\tabcolsep}{4pt}
    \begin{tabular}{@{}l L{5.9cm} L{3.6cm}@{}}
        \toprule
        \textbf{Controller} & \textbf{Decision rule per epoch} & \textbf{Key parameters} \\
        \midrule
        \multicolumn{3}{@{\hfill\,}c@{\,\hfill\hfill}}{\textbf{Placement-only}: choose the node for each stage} \\ \hline
        Round-robin & next feasible node, cyclically & --- \\
        Best-fit & feasible node left with the least spare capacity & --- \\
        Cost-greedy & lowest tier-priced compute plus transfer cost & prices 1/1.6/4/6 (cloud/edge/IoT/space) \\
        MILP & exact min-latency mapping; the only one to relocate a placed stage & SciPy/HiGHS solver \\
        \midrule\multicolumn{3}{@{\hfill\,}c@{\,\hfill\hfill}}{\textbf{Scale-capable}: also set each stage's worker count} \\ \hline
        HPA & scale by queue-per-worker over target; unchanged while that ratio stays within tolerance of 1 & target 0.7 tasks/worker; tolerance 0.1; min 1 worker \\
        KEDA & enough workers that none holds more than the target queue; scale to zero when the queue empties & target 10 tasks/worker; min 0 workers \\
        Joint heuristic & $\pm 1$ worker when queue, age or TTL pressure crosses a threshold; re-plan & queue up 5--18, down 2--4; cooldown 1--2 \\
        Feature-rule & $\pm 1$ worker on a weighted sum of queue, age and TTL pressure & up 1.0, down 0.15; equal weights \\
        Meta-rule & thresholds on the Feature-rule sum with different weights; also decides whether to re-plan and selects the placement strategy & weights 1/0.1/2 (queue/age/TTL) \\
        \bottomrule
    \end{tabular}
\end{table}

\paragraph{\textbf{Protocol settings.}}
The main experiment evaluates all nine controllers on four scenarios under R1 and R2, using ten seeds per configuration. Sensitivity experiments vary one stressor at a time and evaluate six controllers over the same horizon with three seeds each. Findings 1--3 expand on the evaluation by incorporating a replan-schedule ablation, a load sweep, and a placement-stress variant, alongside their corresponding results.

\paragraph{\textbf{Reporting.}}
We report means and standard deviations across all seeds, comparing the two family representatives seed by seed, and interpret ranks together with completion, SLO violation, and p95 latency.

\section{Results}
\label{sec:results}
Table~\ref{tab:main-results} compares the selected placement-only and scale-capable controllers for each scenario--regime pair. In all eight pairs, scale-capable control achieves equal or higher completion, lower SLO violation, and lower p95 latency; the SLO ordering holds in all eighty seed-level comparisons and the completion ordering in all but one seed of the saturated Industrial IoT R1 pair, where mean completion ties at $1.00$. The largest completion gain is in EO under R2, where HPA raises completion from $0.22$ to $0.53$. In Industrial IoT, scaling mainly improves SLO attainment and latency.

\paragraph{\textbf{Finding 1: Placement re-planning has no measurable effect without relocation.}} We fix the scaling policy and recompute placement after every scale event, periodically, or only for newly activated workers. Varying one factor at a time from a common baseline---node count (6, 24 and 48 on a synthesized infrastructure), placement strategy, controller, and regime---over four scenarios (24 configurations $\times$ three schedules $\times$ five seeds, 360 runs), completion differs by less than $10^{-3}$; none of these heuristic schedules relocates an existing stage. Re-planning therefore did not improve incremental placement in any of these configurations. A global MILP marks the boundary: with zero migration delay, it performs approximately $55$ worker-relocation events over the course of each run, increasing completion by $0.045$. This relocation setting is idealized; worker capacity binds in every non-migrating schedule.

\paragraph{\textbf{Finding 2: Scale-capable controllers track the over-provisioned reference as load rises.}} In a separate 480-run R1 capacity sweep, we vary arrival rates from $0.6\times$ to $2.0\times$ and compare completion against an over-provisioned reference that activates all workers at $t=0$. It provides maximum worker capacity but is not a proven completion optimum, because it ignores cost and may constrain placement. HPA, the joint heuristic, and KEDA reach $99$--$100\%$, $97$--$100\%$, and $89$--$100\%$ of this reference; placement-only controllers reach $44$--$100\%$ and degrade with load in every scenario except EO. In all 14 non-saturated scenario--load combinations, every scale-capable controller outperforms placement-only control for every seed. The gap is due to scaling: at fixed capacity, the exact MILP returns the \emph{same} mapping as the min-latency heuristic, and enlarging the synthesized infrastructure to 24 and 48 nodes leaves completion unchanged.

\paragraph{\textbf{Finding 3: Under a placement bottleneck, placement quality affects on-time delivery, but completion-only evaluation hides the effect.}} We construct an EO variant with tighter capacity, a thin edge--cloud link, larger per-task payloads, a $12$\,s deadline, and fixed worker capacity. Cost-greedy completes $102\%$ as many tasks as the best on-time strategy, but delivers only $6\%$ on time versus $47\%$. Completion alone reports a near-tie while hiding an ${\sim}8\times$ on-time-delivery gap. On-time delivery spreads the six strategies by $0.42$ here, $0.12$ with the settings relaxed, and $0.00$ in unmodified EO---placement binds only when it is the bottleneck.

\paragraph{\textbf{Finding 4: The accounting choice changes which controller looks better.}} Under node unavailability, scoring violations over completed tasks alone makes them \emph{decrease} as failures rise in 21 of 24 configurations; counting the unfinished tasks $U$ removes the inversion in all but one. Completion alone also reports no difference between deployed autoscalers: under R2 in Industrial IoT, HPA and KEDA both complete all tasks, yet their completion-aware violation rates are $0.01$ and $0.57$. KEDA scales to zero when its queue empties, so each burst waits until the next control action reactivates workers; under a $10$\,s deadline those waits produce late completions.

\begin{table}[t]
    \centering
    \caption{Main comparison. Each cell reports completion rate\,/\,SLO-violation rate\,/\,p95-latency (s) for the best controller in each family under the selection rule in Section~\ref{sec:metrics}; HPA is reported throughout as the scale-capable representative, being top-ranked in three pairs and within $0.01$ of the top rule-based controller elsewhere. Completion is over all generated tasks, violation over those whose deadline has elapsed. Across ten seeds, std.\ devs.\ are ${\le}0.07$ for completion and ${\le}0.22$ for SLO violation.}
    \label{tab:main-results}
    \footnotesize
    \begin{tabular}{@{}l@{}c@{\hspace{1em}}l@{}r@{}}
        \toprule
        \textbf{Scenario} & \textbf{Reg.} & \hspace{-0.5em}\textbf{Selected scale-capable} & \textbf{Selected placement-only}\hfil\hfil \\
        \midrule
        EO dual stream & R1 & HPA: 0.51/0.34/113.7 & MILP: 0.23/0.75/145.9 \\
        EO dual stream & R2 & HPA: 0.53/0.29/113.4 & Best-fit: 0.22/0.75/133.7 \\
        Driving V2X & R1 & HPA: 0.57/0.35/104.2 & Cost-greedy: 0.49/0.50/114.1 \\
        Driving V2X & R2 & HPA: 0.62/0.26/103.3 & Cost-greedy: 0.53/0.40/110.0 \\
        Split inference & R1 & HPA: 0.86/0.15/72.8 & Best-fit: 0.58/0.54/114.4 \\
        Split inference & R2 & HPA: 1.00/0.00/23.2 & Best-fit: 0.76/0.33/91.1 \\
        Industrial IoT & R1 & HPA: 1.00/0.18/12.2 & Best-fit: 1.00/0.63/24.5 \\
        Industrial IoT & R2 & HPA: 1.00/0.01/8.8 & Best-fit: 0.97/0.78/54.6 \\
        \bottomrule
    \end{tabular}
\end{table}

Targeted stressors also change controller order: contact stress changes 3 of 6 ranks in EO and 4 of 6 in Driving V2X, and load stress 3 of 6 in Industrial IoT, while Split inference remains stable.

Orchestration overhead is negligible at these scales: on the 4--6-node infrastructures of Table~\ref{tab:main-results}, p95 controller decision time stays below $3$\,ms per epoch---under $0.3\%$ of the $1$\,s control period---except for the MILP reference ($8$--$25$\,ms). It does grow steeply with infrastructure size: on the 48-node re-planning variant, per-epoch routing raises even HPA's p95 to $181$\,ms.

For the capacity-bound regimes studied, these results suggest a design guideline: provision elastic capacity first; re-planning and placement optimization paid off only once worker capacity had ceased to be the bottleneck.

\paragraph{\textbf{Limitations and threats to validity.}}
ContinuumBench is simulation-only. Its contact, buffering, and startup models lack hardware validation, so results are relative comparisons; a Kubernetes-backed harness is the next step. Results are also conditional on ECLYPSE, which models feasibility, residual resources, and routing but omits transport contention, container scheduling, and handover dynamics. The infrastructures are hand-built graphs with 4--6 nodes (up to 48 in the replan ablation), and the scenarios and R2 calibration profiles are representative rather than externally validated.

Quantizing sub-second per-hop delays to the $1$\,s control epoch establishes a baseline latency floor equal to the longest path length (${\sim}6$\,s in Industrial IoT, against a $10$\,s deadline). Because this floor applies identically across controllers and compresses observed separations rather than creating them, absolute latency inflation does not compromise relative comparisons.

Finally, the evaluated regimes are capacity-bound (Findings 1--2), and because the heuristics never relocate, only the migrating MILP optimizer tests re-planning.

\section{Conclusions and Future Work}
By isolating individual control levers and exposing evaluation assumptions, ContinuumBench shows that elastic capacity drives completion in the regimes evaluated here, and that once capacity suffices, the choice of autoscaling policy decides how much of that work arrives on time. It also shows that the accounting choice changes the reported result: completion-only and completion-aware scoring can rank the same controllers differently.

Four natural extensions follow. First, introducing an explicit cost model will test whether anticipatory control becomes advantageous when over-provisioning is penalized. Second, the controller set covers production policies and an exact optimizer but no learning-based methods; adding them is the obvious next step, and ContinuumBench already logs the per-epoch observations, actions, and seeds they need. Third, scaling to thousand-node networks requires addressing per-epoch routing overhead via path caching, candidate pruning, and replacing the exact MILP reference. Finally, evaluating synthetic topologies (e.g., Erd\H{o}s--R\'enyi, Barab\'asi--Albert) will test whether our capacity-bound findings hold under sparser connectivity.

\begin{credits}
\subsubsection{\ackname}
This research was partially supported by the FIS2 Grant from the Italian Ministry of University and Research (Grant ID: FIS2023-03382).
\end{credits}

%
%
%
\bibliographystyle{splncs04}
\bibliography{references}
\end{document}